\documentclass[preprint,1p]{elsarticle}
\usepackage{graphicx}
\usepackage{amsmath}
\usepackage{natbib}
\usepackage[usenames]{color}
\usepackage{soul}

\usepackage[normalem]{ulem}

\usepackage[utf8]{inputenc}

\usepackage{amsfonts}
\usepackage{dcolumn}        
\usepackage{amssymb}
\usepackage{bm,amssymb}

\usepackage{bm,fancybox}                	     

\begin{document}

\newcommand{\bea}{\begin{eqnarray}}
\newcommand{\eea}{  \end{eqnarray}}
\newcommand{\bit}{\begin{itemize}}
\newcommand{\eit}{  \end{itemize}}

\newcommand{\be}{\begin{equation}}
\newcommand{\ee}{\end{equation}}
\newcommand{\ra}{\rangle}
\newcommand{\la}{\langle}
\newcommand{\U}{\widetilde{U}}

\def\bra#1{{\langle#1|}}
\def\ket#1{{|#1\rangle}}
\def\bracket#1#2{{\langle#1|#2\rangle}}
\def\inner#1#2{{\langle#1|#2\rangle}}
\def\expect#1{{\langle#1\rangle}}
\def\e{{\rm e}}
\def\proj{{\hat{\cal P}}}
\def\tr{{\rm Tr}}
\def\H{{\hat H}}
\def\Hdag{{\hat H}^\dagger}
\def\Lop{{\cal L}}
\def\Ehat{{\hat E}}
\def\Edag{{\hat E}^\dagger}
\def\Shat{\hat{S}}
\def\Sdag{{\hat S}^\dagger}
\def\Ahat{{\hat A}}
\def\Adag{{\hat A}^\dagger}
\def\U{{\hat U}}
\def\Udag{{\hat U}^\dagger}
\def\Zhat{{\hat Z}}
\def\Phat{{\hat P}}
\def\Op{{\hat O}}
\def\id{{\hat I}}
\def\x{{\hat x}}
\def\P{{\hat P}}
\def\Px{\proj_x}
\def\Pr{\proj_{R}}
\def\Pl{\proj_{L}}


\title{Exploring quantum localization with machine learning}
\author[add3]{Javier Montes}
 \ead{jmontes.3@alumni.unav.es}

\author[add2]{Lenoardo Ermann}
 \ead{leonardoermann@cnea.gob.ar}

\author[add2]{Alejandro M.F. Rivas}
 \ead{rivas@tandar.cnea.gov.ar}

\author[add1]{Florentino Borondo}
 \ead{f.borondo@uam.es}

\author[add2]{Gabriel G. Carlo}
\ead{gabrielcarlo@cnea.gob.ar}
 
 \address[add1]{Departamento de Qu\'imica, 
 Universidad Aut\'onoma de Madrid,
 Cantoblanco, 28049--Madrid, Spain}

 \address[add2]{Comisi\'on Nacional de Energ\'ia At\'omica, CONICET, Departamento de F\'isica, 
 Av.~del Libertador 8250, 1429 Buenos Aires, Argentina}

 \address[add3]{Department of Statistics, Universidad Carlos III de Madrid, Spain}
 
\date{\today}

\begin{abstract}
We introduce an efficient neural network (NN) architecture for classifying wave functions in terms of their localization. Our approach integrates a versatile quantum phase space parametrization leading to a custom "quantum" NN, with the pattern recognition capabilities of a modified convolutional model. This design accepts wave functions of any dimension as inputs and makes accurate predictions at an affordable computational cost. This scalability becomes crucial to explore the localization rate at the semiclassical limit, a long standing question in the quantum scattering field. Moreover, the physical meaning built in the model allows for the interpretation of the learning process.
\end{abstract}
\maketitle

\section{Introduction}
 \label{sec:intro}

Determining the localization degree and morphology of resonances in partially open quantum systems (crucial to the study of resonant cavities and scattering in general) has been a long awaited objective attacked from many sides \cite{Blumel1,Blumel2,Mendez,art0,Dietz,Novaes,Cao,RecentExp}. Great advances of 
undoubted relevance are contained in these works, but the precise goal has proven elusive until apparently very recent times, when new promising tools, conjectures and theories have been developed \cite{HusimisNew,Ketz1,Ketz2,Ketz3,Ketz4,Ketz5,Ketz6,our1,our2,our3}. 
But these results need more systematic tests and thorough explorations of large 
sets of resonances in the semiclassical limit. This is a formidable task since not 
only the calculation of the eigenfunctions is needed, but also the computation of 
their quantum phase space representations. Moreover, one needs to compute some 
suitable localization measure reasonably invariant on the representation used in order to have an estimation of the localization degree. The advent of Machine Learning (ML) gives us the opportunity to overcome this last barrier. 

The use of ML techniques has extended to the vast majority of research areas nowadays. The physical sciences are no exception \cite{ML1}, and successful applications can be found in quantum entanglement and complexity \cite{ML2,ML3,ML4}, a subject closely related to this work. On the other hand, NNs (a ML technique) have demonstrated extremely powerful in image processing, being this a key feature we exploit \cite{AlexNet}.   

In this paper we focus on the localization properties of the resonances associated to the open quantum tribaker map, which is a paradigmatic model originally proposed in the quantum chaos area but that has many applications in quantum optics, quantum computation and scattering processes in general \cite{art1,art2,art3}. We use the so called norm ratio \cite{art4,art5} as the localization measure labelling the data in order to train a NN. This NN is composed of two parts, one dedicated to quantum phase space sensing by a suitable adaptation of the Husimi distribution, and the other devoted to recognize localization as an image feature. Thanks to thorough tests of our model, we have found that it would be possible to extend the estimation of the localization degree to high Hilbert space dimensions, which in our setting amounts to reaching the semiclassical limit. Moreover, our ML model unveils localization properties of resonances showing how short periodic orbits (POs) could be hidden in the set of long lived resonances thanks to specially suited learning features built in our NN.

The organization of this paper is as follows: 
in Sec.~\ref{sec:SystemandMeasure} we describe the (partially) open tribaker map 
and the localization measure used. In Sec.~\ref{sec:NNarchitecture} we describe the architecture of our NN. In Sec.~\ref{sec:DatasetTraining} we show how we handle the dataset and train our ML model. Finally, in Sec.~\ref{sec:Results} we annalyze its performance. Some relevant points are outlined in Sec.~\ref{sec:Conclusions}.

\section{System and measure}
 \label{sec:SystemandMeasure}

We here give a brief description of (partially open) maps that are an invaluable resource in  classical and quantum chaos \cite{Ozorio 1994,Hannay 1980,Espositi 2005}. 
The main invariant structure in phase space associated to open maps on the 2-torus is the fractal repeller, which is defined as the intersection of the forward and backwards trapped sets, which in turn are constructed by the nonescaping trajectories in the past or future. If the opening is not complete there is a finite reflectivity (in our example we take a constant function $R \in (0:1)$), the trajectories have a variable intensity, and the now multifractal invariant extends over all the phase space.

Throughout this paper we consider the tribaker map, which in its closed 
and classical version is 
\begin{equation}
\mathcal B(q,p)=\left\{
  \begin{array}{ll}
  (3q,p/3) & \mbox{if } 0\leq q<1/3 \\
  (3q-1,(p+1)/3) & \mbox{if } 1/3\leq q<2/3\\
  (3q-2,(p+2)/3) & \mbox{if } 2/3\leq q<1\\
  \end{array}\right.
\label{classicaltribaker}
\end{equation}
We take the one with an opening in the region $1/3< q<2/3$, i.e. all trajectories passing through it will modify their intensity following the previously mentioned reflectivity function.

In general, the quantization of maps amounts to taking 
$\bracket{q+1}{\psi}\:=\:e^{i 2 \pi \chi_q}\bracket{q}{\psi}$, and
$\bracket{p+1}{\psi}\:=\:e^{i 2 \pi \chi_p}\bracket{p}{\psi}$, with $\bracket{q}{\psi}$ and $\bracket{p}{\psi}$ the wave functions in position and momentum basis, respectively ($\chi_q$, $\chi_p \in [0,1)$). Consequently, the dimension $N$ of the Hilbert space satisfies $N=(2 \pi \hbar)^{-1}$, and $N \rightarrow \infty$ means going to the semiclassical limit. The evolution operator is a
$N\times N$ matrix, while the position and momentum eigenstates are given by
$\ket{q_j}\:=\:\ket{(j+\chi_q)/N}$ and $\ket{p_j}\:=\:\ket{(j+\chi_p)/N}$ with
$j\in\{0,\ldots, N-1\}$, and 
$\bracket{p_k}{q_j}\:=\: \frac{1}{\sqrt{N}} e^{-2i\pi(j+\chi_q)(k+\chi_p)/N} \: \equiv \:
(G^{\chi_q, \chi_p}_N)$. When the quantum map is (partially) open the corresponding operator is non-unitary, having 
$N$ right eigenvectors $|\Psi^R_j\ra$ and $N$ left ones $\la \Psi_j^L|$ 
for each resonance (eigenvalue) $z_j$, 
with $\la \Psi_j^L|\Psi^R_k\ra=\delta_{jk}$ and 
$\la \Psi_j^R|\Psi^R_j\ra=\la \Psi_j^L|\Psi^L_j\ra$ being the norm.
The quantum tribaker map (we take antiperiodic boundary conditions 
$\chi_q=\chi_p=1/2$ in order to preserve its classical symmetries) in position representation is \cite{Saraceno1,Saraceno2}
\begin{equation}\label{quantumbaker}
 U^{\mathcal{B}}=G_{N}^{-1} \left(\begin{array}{ccc}
  G_{N/3} & 0 & 0\\
  0 & G_{N/3} & 0\\
  0 & 0 & G_{N/3}\\
  \end{array} \right).
\end{equation}
The partially open map is recovered when we apply the projector  
\begin{equation}\label{partialprojector}
 P=\left(\begin{array}{ccc}
  1_{N/3} & 0 & 0\\
  0 & \sqrt{R} \; 1_{N/3} & 0\\
  0 & 0 & 1_{N/3}\\
  \end{array} \right),
\end{equation}
 to the evolution given by Eq.~(\ref{quantumbaker}).

We define a symmetrical operator $\hat{h}_j$ which takes into account both  
the right $\vert \Psi^R_j\rangle$ and left $\langle \Psi^L_j\vert$ eigenstates \cite{art2,art3}
\begin{equation}\label{eq.hdef}
 \hat{h}_j=\frac{\vert \Psi^R_j\rangle\langle \Psi^L_j\vert}{\langle \Psi^L_j\vert \Psi^R_j\rangle}.
\end{equation}
It is worth noticing that adding all these symmetrical operators multiplied by the corresponding eigenvalue $z_j$ is equivalent to the spectral decomposition of the non-unitary operator. Since the right (left) eigenstates have support on the backwards (forward) trapped sets, $\hat{h}_j$ are concentrated on the classical repeller. If we take expectation values on coherent states $|q,p\rangle$, it follows that $h_j(q,p) = \vert\langle q,p\vert \hat{h}_j\vert q,p\rangle\vert \propto 
\sqrt{{\cal H}^R_j (q,p) {\cal H}^L_j (q,p)}$, 
with ${\cal H}^{R,L}_j$ the Husimi distributions of the $R, L$ eigenstates 
(which in the closed case coincide and are just the usual Husimi distributions). We call $h_j(q,p)$ the {\em LR representation} of the resonances (the modulus is taken since these are complex functions). 

The measure of localization that we have selected in order to label the resonances  considers a coherent state as the most localized distribution and by comparing phase space norms reflects the departure from it.
This is the norm ratio $\mu$ defined by \cite{art4}
\begin{equation}\label{eq.normratio}
 \mu(\tilde{h}_i)=\left( \frac{{\|\tilde{h}_i\|}_1/{\|\tilde{h}_i\|}_2}{{\|\rho_c\|}_1/{\|\rho_c\|}_2} \right)^2.
\end{equation}
A coherent state at $(q,p)$ provides the normalization factor $\rho_c= \vert q,p \rangle \langle q,p \vert$ by means of the phase space norm 
\begin{equation}\label{eq.phasenorm}
 {\|\tilde{h}_i\|}_\gamma=\left( \int_{{\cal{T}}^2} {\tilde{h}_i(q,p)^\gamma dq dp } \right)^{1/\gamma}.
\end{equation}
The norm ratio does not depend on the $h$ normalization, reaching a  
minimum of $1$ for a maximally localized distribution (a coherent state) 
and a maximum of $N/2$ for the uniform distribution. 

\section{NN architecture}
 \label{sec:NNarchitecture}
\subsection{Custom layer optimized for wave function convolution}

\begin{figure}[ht]
\centering
\includegraphics[width=\linewidth]{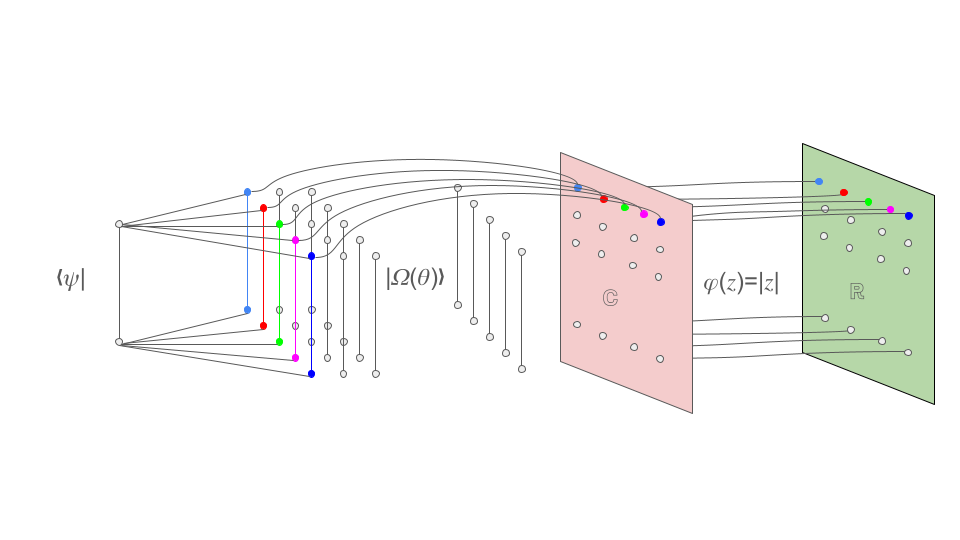}
\caption{Schematic representation of the NN. The input to the network is the bra \(\langle\psi|\), and it undergoes a series of convolutions with parametrized functions \(\Omega(\theta)\), each represented by a different color. The convolution gives a corresponding point in the output "image" (point colors match \(\Omega(\theta)\) colors). The image is processed through an activation function, resulting in the final image output.}
\label{fig1}
\end{figure}

Our architecture introduces a streamlined approach to processing quantum wave functions, starting with a custom convolution layer specifically optimized for this task (see Fig. \ref{fig1}). This layer is designed to convolve the quantum wave function, denoted by $\langle \psi |$, with a parametrizable kernel represented by $\ket{\Omega}$. The convolution transforms the quantum mechanical data into a format that is compatible with traditional deep learning techniques. Regarding the parametrizable kernels we can explicitly define them as \(\langle x | \Omega(\theta)\rangle\), where \(x\) represents the spatial coordinates  (i.e. the position basis) and \(\theta\) denotes a set of parameters. These kernels are generated dynamically taking into account the dimension of the wave function, and the specific parameters \(\theta\) are tailored to optimize the convolution process. The parametrizable nature of the kernels allows for a flexible adjustment to the morphology of each wave function, ensuring that the convolution accurately captures the essential features of the quantum state.

The convolution operation in this custom layer is given by
\begin{equation}
\langle \psi | \Omega \rangle_{ab} = \int \psi(x) \cdot \Omega_{ab}(x;\theta) dx
\end{equation}
where $\langle \psi | \Omega \rangle_{ab}$ represents the convolved wave function, capturing the localization features in the parameter space. We denote $\psi$ for brevity, but stands for both $\Psi^L$ and $\Psi^R$, the corresponding convolutions are then combined (in the $R=1$ case they coincide). This process effectively translates the quantum information into a set of "images", each bearing the degree of phase space localization of each wave function. The function \(\Omega(x; \theta)\) is arbitrary but its choice depends on the specific problem at hand, as well as the information sought from the wave functions. We have taken \(\Omega\) as coherent states \(\alpha\) (in position representation) with parameters \(\theta\) given by their positions \(\{p_0, q_0\}\) in phase space, which are adjustable during training. Upon convolving \(\Omega\) with a wave function \(\psi\), and subsequently applying the modulus operation via the activation function \(\varphi(z) = \|z\|\), the outcome is effectively transformed into the LR representation $h_j(q,p)$, allowing to physically interpret the convolution results.

\subsection{Integration with a modified AlexNet}

Following the convolution with the custom layer, our architecture employs a modified version of the AlexNet deep learning model \cite{AlexNet} adapted to handle the structure of our images. The AlexNet model, originally designed for RGB images, is reconfigured to accept a single-channel input, corresponding to the output of the custom convolution layer. The modified AlexNet consists of several convolutional layers, pooling layers, and fully connected layers, structured as follows:
\begin{itemize}
\item Convolutional layers extract features from our images, using filters to capture patterns signaling localization in the phase space.
\item Pooling layers reduce the dimensionality of the feature maps, enhancing the network's ability to generalize from the data by focusing on dominant features.
\item Fully connected layers consolidate the extracted features into a form suitable for classification, ending with a final layer that informs the  (non-)localized nature of the wavefunction.
\end{itemize}

\begin{figure}[ht]
\centering
\includegraphics[width=\linewidth]{"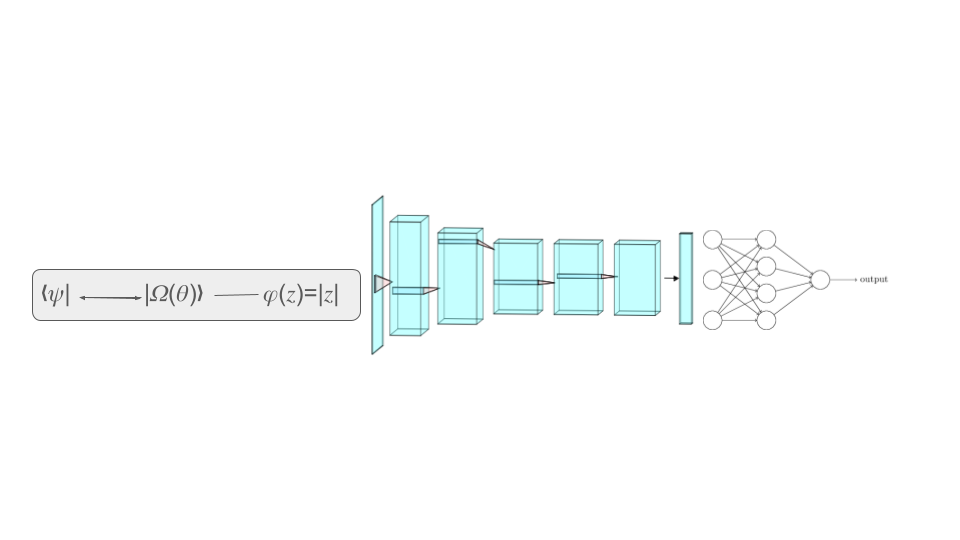"}
\caption{NN architecture. The network input,\(\langle\psi|\), undergoes convolution with the parametrized function \(\ket{\Omega}\), followed by an activation function. This is fed to a modified single input channel AlexNet, concluding with a perceptron layer consisting of a single output neuron.}
\label{fig2}
\end{figure}

The entire network (see Fig. \ref{fig2}) operates in a supervised training regime, where the input wave functions are labeled according to their localization properties in quantum phase space. The training process aims to minimize a loss function that quantifies the difference between the predicted and actual labels, employing backpropagation and gradient descent algorithms to adjust the parameters of both the custom convolution layer and the AlexNet model. 

\section{Dataset handling and NN training}
 \label{sec:DatasetTraining}

We construct the dataset in order to train our NN starting with a Hilbert space of dimension $N = 243$. For each value of $N$, we calculate the resonances of the quantum system, resulting in a total of $N$ left and $N$ right eigenstates, which in the closed system ($R=1$) amounts to just $N$ eigenstates. Following this we compute the norm ratio $\mu(\psi_i)$ for each one of them, considering just the $20$ wave functions with the highest norm ratios and the $20$ ones with the lowest values. These selected wave functions are then labeled accordingly: a label $0$ is assigned to resonances with the lowest norm ratios (non-localized), and $1$ is assigned to those with the highest values (localized). After the initial iteration, we grow $N$ by 3 and repeat the entire process. This iterative procedure is conducted $50$ times, increasing $N$ up to $N_{max}=243 + 3 \times 50$ to give a total of $2 \times 40 \times 50$ labeled left and right wave functions (again, in the closed case they are just half this number).

To compound the wave functions into a unified dataset, we construct the matrix \(\Psi_{\mu\nu}\) (we take the closed case as an example)
\[
\Psi_{\mu\nu} = 
\left[
\begin{array}{cccc}
\psi_1(x_1) & \cdots & \psi_{k}(x_1) & \psi_{40\times50}(x_1) \\
\vdots & \ddots & \vdots & \vdots \\
\psi_1(x_{N_1}) & \cdots & \psi_{k}(x_{N_k}) & \psi_{40\times50}(x_{N_{k}}) \\
0 & \cdots & 0 & \psi_{40\times50}(x_{N_{k+1}}) \\
\vdots & \ddots & \vdots & \vdots \\
0 & \cdots & 0 & \psi_{40\times50}(x_{N_{\text{max}}}) \\
N_1 & \cdots & N_k & N_{\text{max}} \\
\end{array}
\right],
\]
where the \(\nu\)-th column corresponds to the \(\psi_\nu(x)\) wave function, and each row \(\mu\) represents the coefficient of the projection of the eigenstate \(\psi_\nu\) onto the position basis element \(x_\mu\), such that \(\langle x_\mu|\psi_\nu\rangle = \psi_\nu(x_\mu)\). To accommodate wave functions of varying dimensions within the same matrix, we append zeros to the entries beyond the dimension \(N_\nu\) of each wave function \(\psi_\nu\), thereby standardizing the column lengths. The dimension \(N_\nu\) of the wave function \(\psi_\nu\) is explicitly recorded in the last row of the matrix.

Summarizing, the dataset for our study is constructed from a series of quantum wave functions, each encapsulated within a tensor, denoted as \(\Psi_{\mu\nu}\). The labels associated with each wave function are binary, determined by a calculated norm ratio, with a value of 1 indicating a non-localized state and 0 denoting a localized one. Formally, the dataset \(\mathcal{D}\) comprises a collection of tuples, where each tuple \((\Psi_{\mu\nu}, y)\) contains the tensor representation of a wave function and its associated binary label \(y\). This structure ensures that each wave function is suitably prepared for processing by the NN, adhering to the required input specifications.

To ensure a robust training and validation process, the dataset \(\mathcal{D}\) is systematically divided into training (\(\mathcal{D}_{\text{train}}\)) and validation (\(\mathcal{D}_{\text{val}}\)) subsets. This division is executed randomly to maintain a representative distribution of wave functions across both subsets, thereby facilitating a comprehensive training regime and an unbiased validation process. Typically, 80\% of the dataset is designated for training, with the remaining 20\% allocated for validation purposes. Additionally, the training and validation data are organized into batches, with a standard batch size of 32, to enhance the efficiency and effectiveness of the training phase.

Our ML model, designed to classify quantum wave functions based on their localization features, has an initialization which sets weights and biases to values that are optimized for learning, focusing on compatibility with ReLU activation functions. The optimization of the NN parameters is conducted using the Adam optimization algorithm, targeting the minimization of cross-entropy loss between the model predictions and the actual binary labels.

The training protocol encompasses multiple epochs, each representing a full iteration through the \(\mathcal{D}_{\text{train}}\) subset. Within each epoch, the model performs forward propagation to compute predictions, evaluates the loss to determine the accuracy of these predictions against the actual labels, and does backpropagation to adjust its parameters based on the gradient of the loss. To warrant the generalization capability of our model and mitigate overfitting, its performance is regularly evaluated on the \(\mathcal{D}_{\text{val}}\) subset, allowing for iterative refinement of the training strategy based on empirical validation results.

\section{Results and performance of the ML model}
 \label{sec:Results}

\begin{figure}[ht]
\centering
\includegraphics[width=0.8\textwidth]{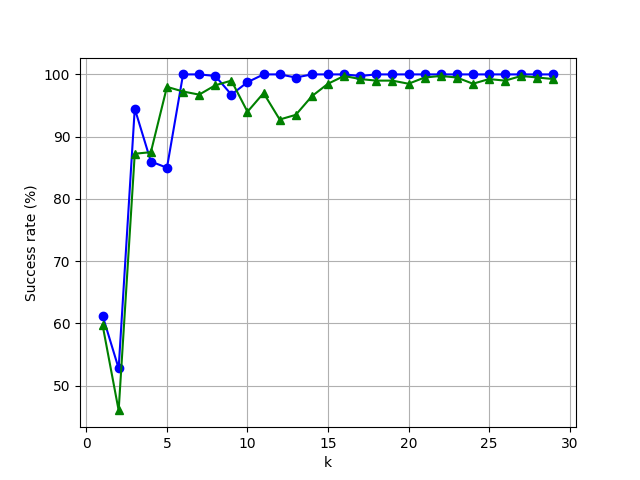}
\hfill
\caption{(Color online) Success rate as a function of $k$. The partially open system ($R=0.07$) is represented in blue (black), while the closed system in green (gray).}
\label{fig3}
\end{figure}

\begin{figure}[ht]
\centering
\includegraphics[width=0.8\textwidth]{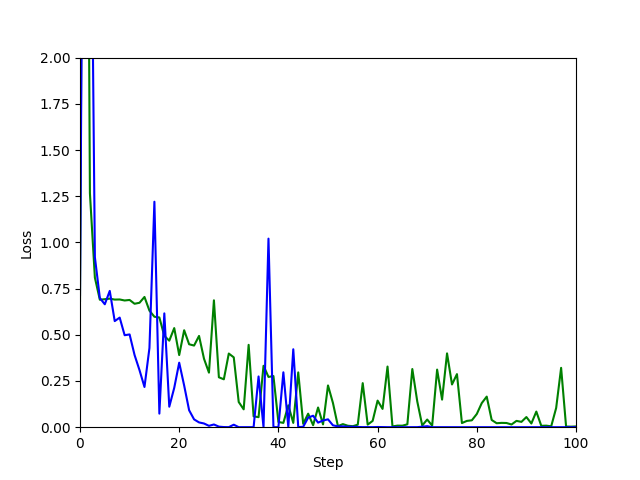}
\hfill
\caption{(Color online) Loss function versus training steps using the entire dataset ($k=50$). The partially open system is represented in blue (black) while the closed one in green (gray).}
\label{fig4}
\end{figure}

The first thing to notice regarding the performance of our NN is that within a range of up to 5 training epochs (each epoch represents a complete iteration through the dataset) an accurate classification of 100\% of the wave functions has been obtained. This indicates rapid convergence towards an optimal solution, demonstrating the effectiveness of the network architecture and the adequacy of the dataset for the proposed classification task. Also, the NN demonstrates scalability for wave functions of varying dimension \(N\), being capable of classifying both higher and lower dimensional wave functions. And finally, the NN learns with significantly reduced datasets, becoming applicable in scenarios with limited data. In order to better grasp the performance of our model, in Fig. \ref{fig3} we show the rate of success over the validation dataset as a function of the number $k$ of subsets of different $N$ used in the traning process. Here, $k=1$ corresponds to the first $40$ wave functions used for training, $k=2$ to $80$ (we use both the sets for $N=243$ and $N=246$) and so on up to $k=30$. 
It can be clearly seen that the partially open situation is even more amenable to the localization exploration of our NN than the already very good closed scenario, since it reaches an almost perfect rate at just $k=6$, compared to $k=16$ in the latter case. 
When we fully train our model using $k=50$ and 5 epochs, with the dataset being traversed in each epoch, we find the behavior of the loss function shown in Fig. \ref{fig4}. It becomes clear that the learning process completely stabilizes at around the 50th step (a step refers to an iteration within the training algorithm) in the partially open case, while in the closed one there are still fluctuations (though small), up to the very end. This is another way to see the suitability of our model for exploring localization in partially open systems.

\begin{figure}[ht]
\centering
\includegraphics[width=0.8\textwidth]{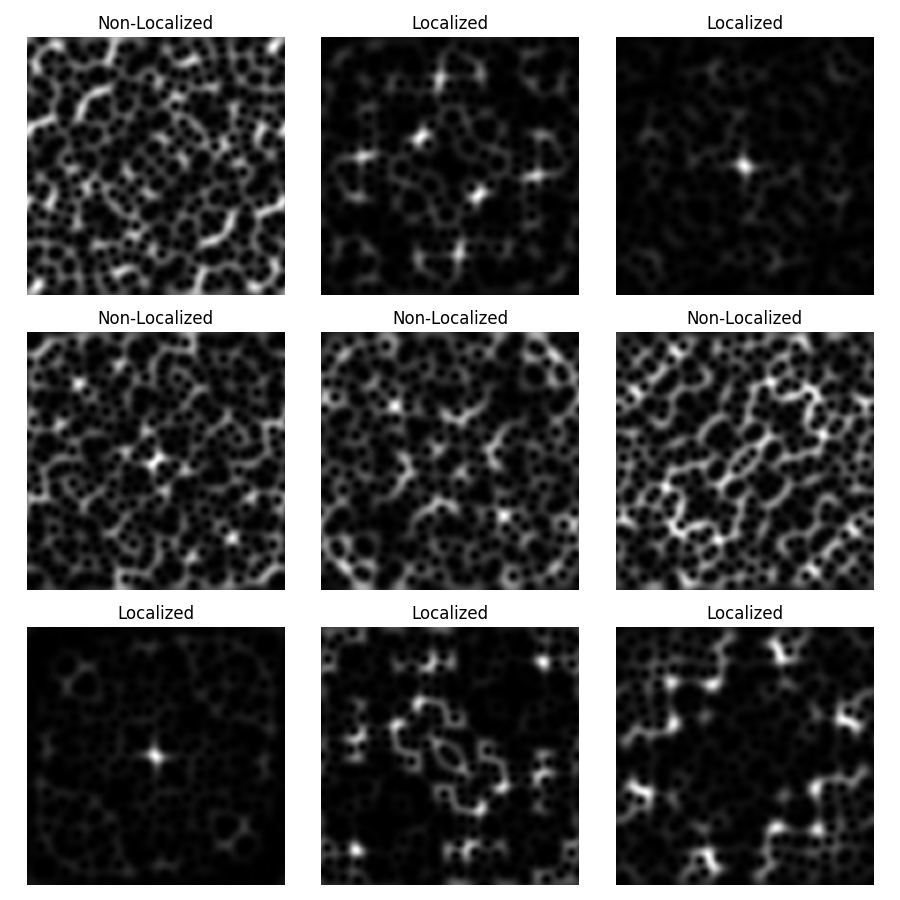}
\caption{Husimi distributions (closed case) of 9 randomly selected wave functions from the validation dataset, together with their classification in terms of localization.}
\label{fig5}
\end{figure}

Now we go deeper into the interpretation of the results we have obtained. For the closed case, Fig. \ref{fig5} shows the Husimi distributions (as $R=1$ the LR representations reduce to them) of 9 randomly selected eigenfunctions from the validation dataset, all of which have been correctly classified. It is observable that the localization patterns differ across each wave function, as do the non-localization features. Despite these striking differences, the network has successfully classified them. Then, how does this classification work?

\begin{figure}[ht]
\centering
\includegraphics[width=0.8\textwidth]{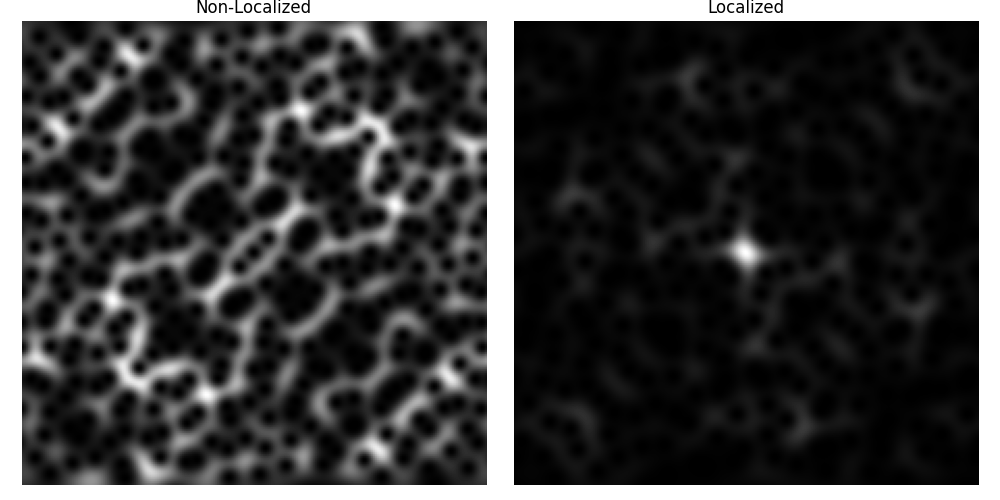}
\hfill
\caption{Husimi distributions selected to illustrate the convolutional filters action. On the left we find the non-localized case, while on the right the localized one.}
\label{fig6}
\end{figure}

To classify wave functions, any given input will activate certain layers and neurons within the NN to finally provide with a binary outcome, 0 or 1. In particular, we integrate a modified AlexNet where these layers are convolutional filters that progressively transform the image generated in our "quantum" NN to arrive at this binary classification. Although the NNs learning process is not directly interpretable in general, examining how kernels deform the input can offer an approximate understanding of it. For that purpose we have selected two wave functions from those shown in Fig. \ref{fig5}, the localized 3rd one (from left to right and top to bottom) and the non-localized 6th one (see Fig. \ref{fig6} for clarity). 

\begin{figure}[ht]
\centering
\includegraphics[width=0.45\textwidth]{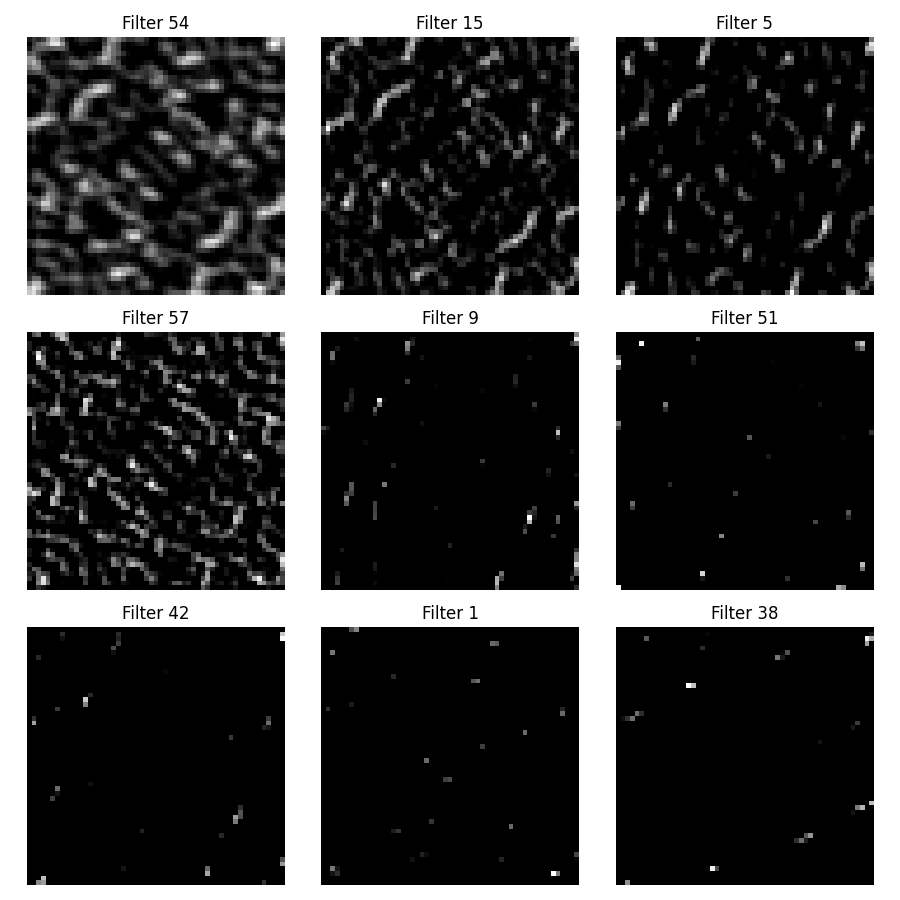}
\hfill
\includegraphics[width=0.45\textwidth]{Figure_7.png}
\caption{Effect of the top 9 activating convolutional filters in the first layer for the cases of Fig. \ref{fig6}. In the right panels (localized example), the short PO scar identified by our NN is marked by means of blue empty circles.}
\label{fig7}
\end{figure}

\begin{figure}[ht]
\centering
\includegraphics[width=0.45\textwidth]{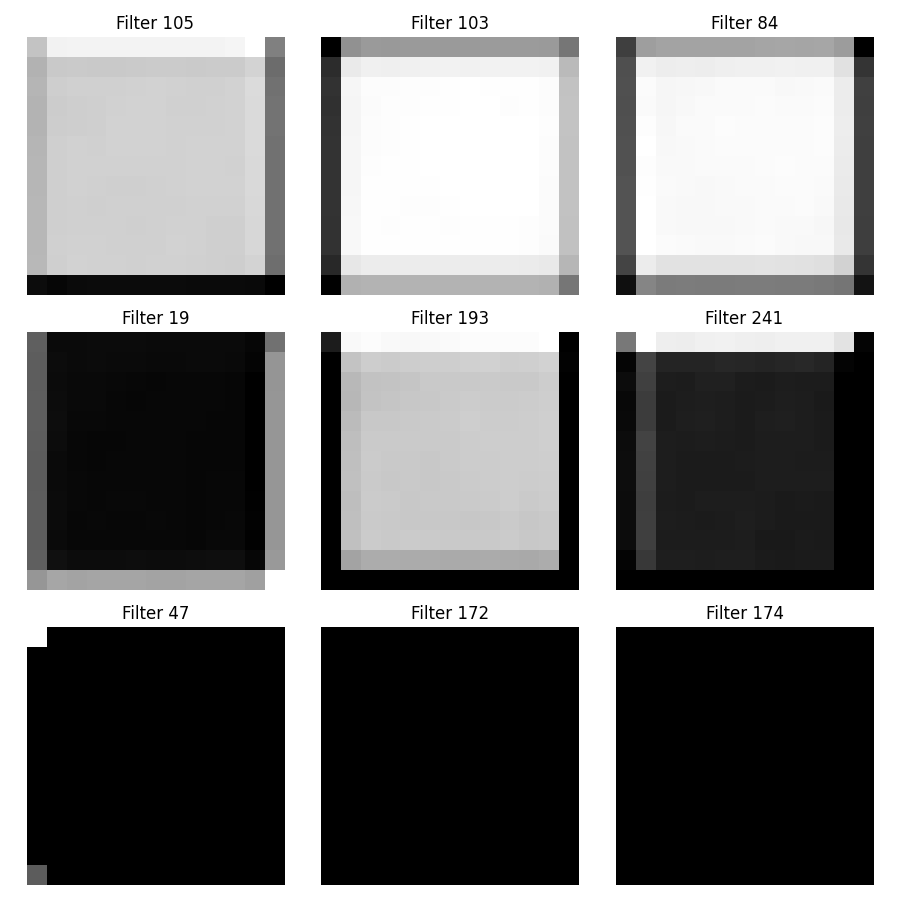}
\hfill
\includegraphics[width=0.45\textwidth]{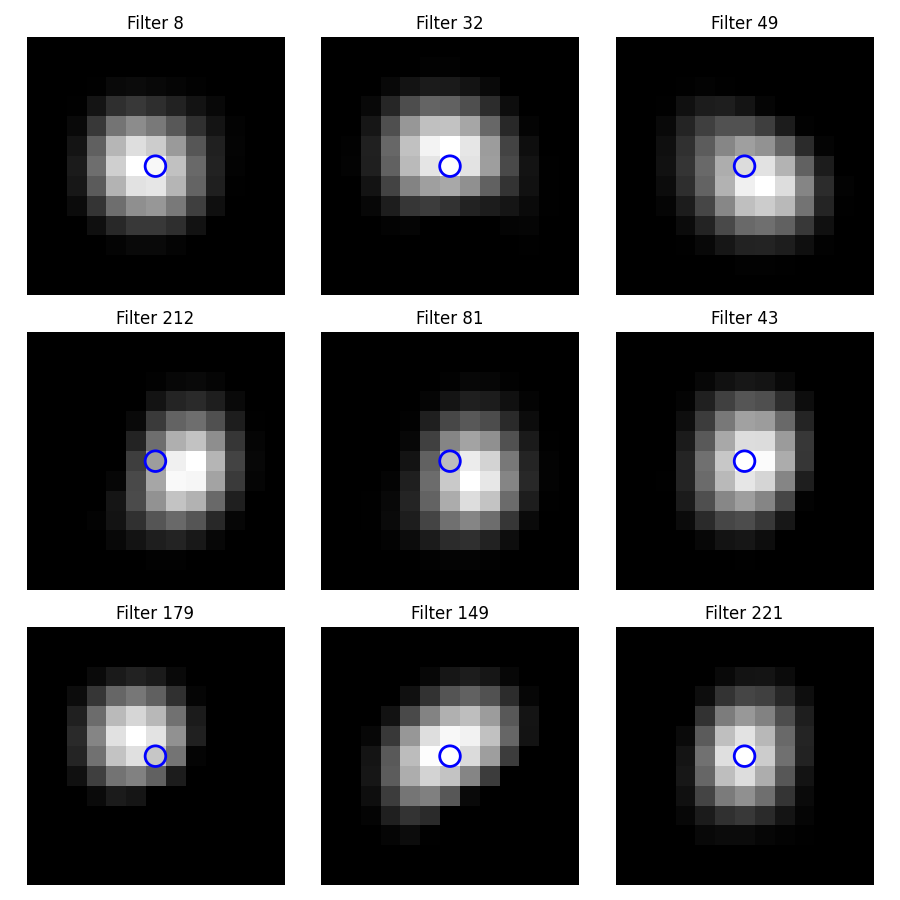}
\caption{Effect of the top 9 activating convolutional filters in the last layer for the cases of Fig. \ref{fig6}. In the right panels (localized example), the short PO scar identified by our NN is marked by means of blue empty circles.}
\label{fig8}
\end{figure}

In Fig. \ref{fig7} we display the input images after passing through the top 9 activating convolutional filters in the first layer of kernels for the cases shown in Fig. \ref{fig6}. We observe that the image suffers a minimal deformation for the few first filters, although it is evident that they progressively make the distributions more uniform. This effect is stronger for the non-localized state, where the filters almost completely wipe out any feature at the end of the process.    
This can be seen with the help of Fig. \ref{fig8} where we show the last (5th) layer. In the localized case we observe that from the first to the last layer, all filters focus on a short PO of period 1 that strongly scars the eigenfunction. In the non-localized example the filtered distribution ends up being uniformly extended over the whole phase space.

\begin{figure}[ht]\centering
\includegraphics[width=0.8\textwidth]{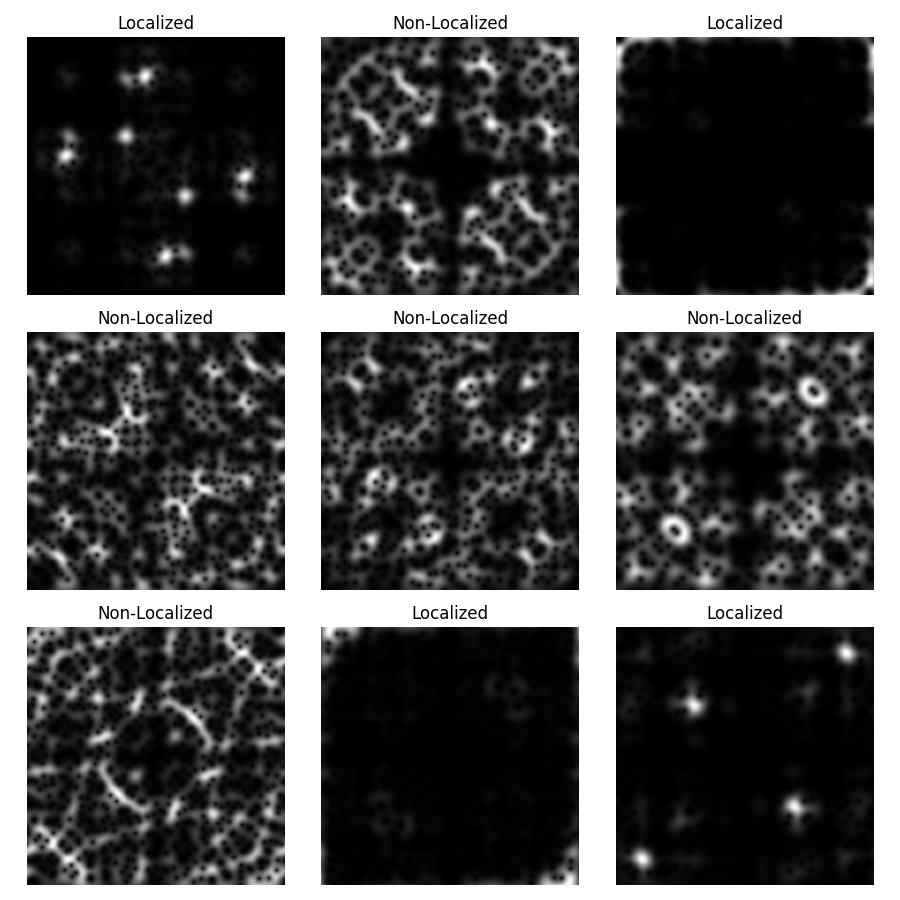}
\caption{LR representations (partially open case, $R=0.07$) of 9 randomly selected wave functions from the validation dataset, together with their classification in terms of localization.}
\label{fig9}
\end{figure}

\begin{figure}[ht]
\centering
\includegraphics[width=0.8\textwidth]{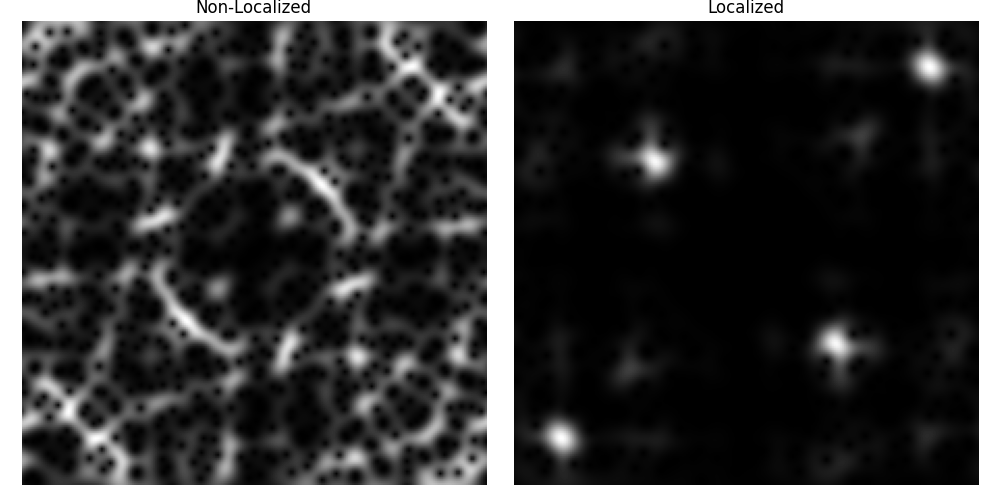}
\hfill
\caption{LR representations selected to illustrate the convolutional filters action. On the left we find the non-localized case, while on the right the localized one.}
\label{fig10}
\end{figure}

\begin{figure}[ht]
\centering
\includegraphics[width=0.45\textwidth]{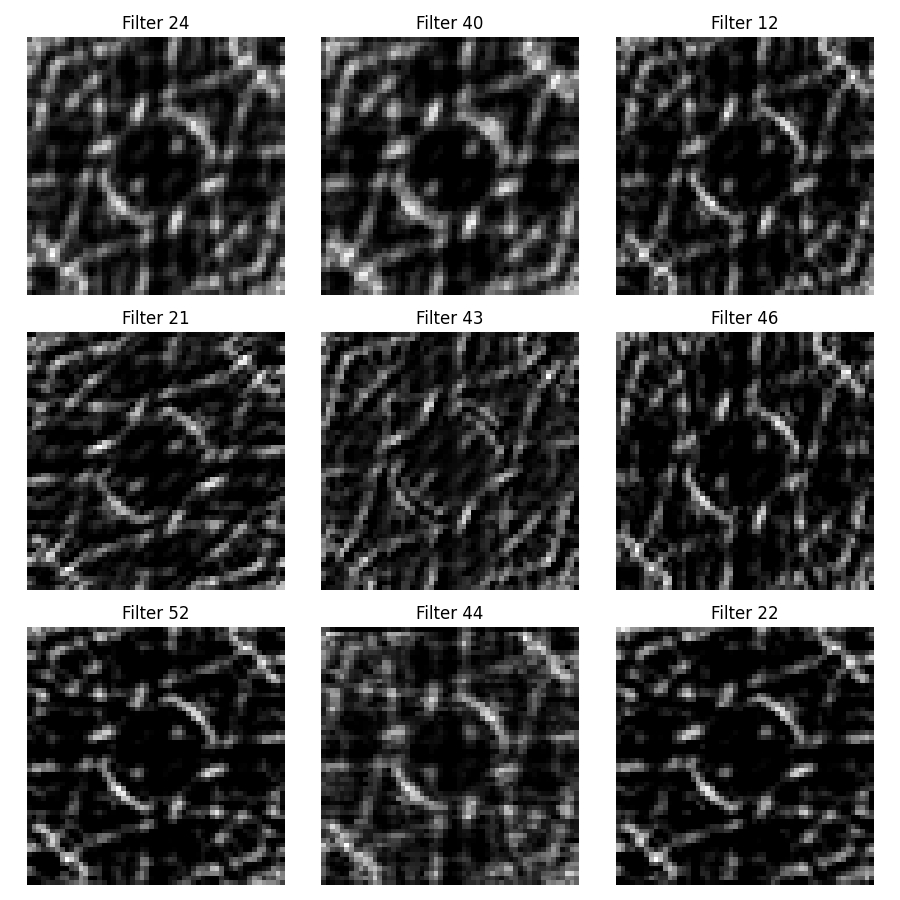}
\hfill
\includegraphics[width=0.45\textwidth]{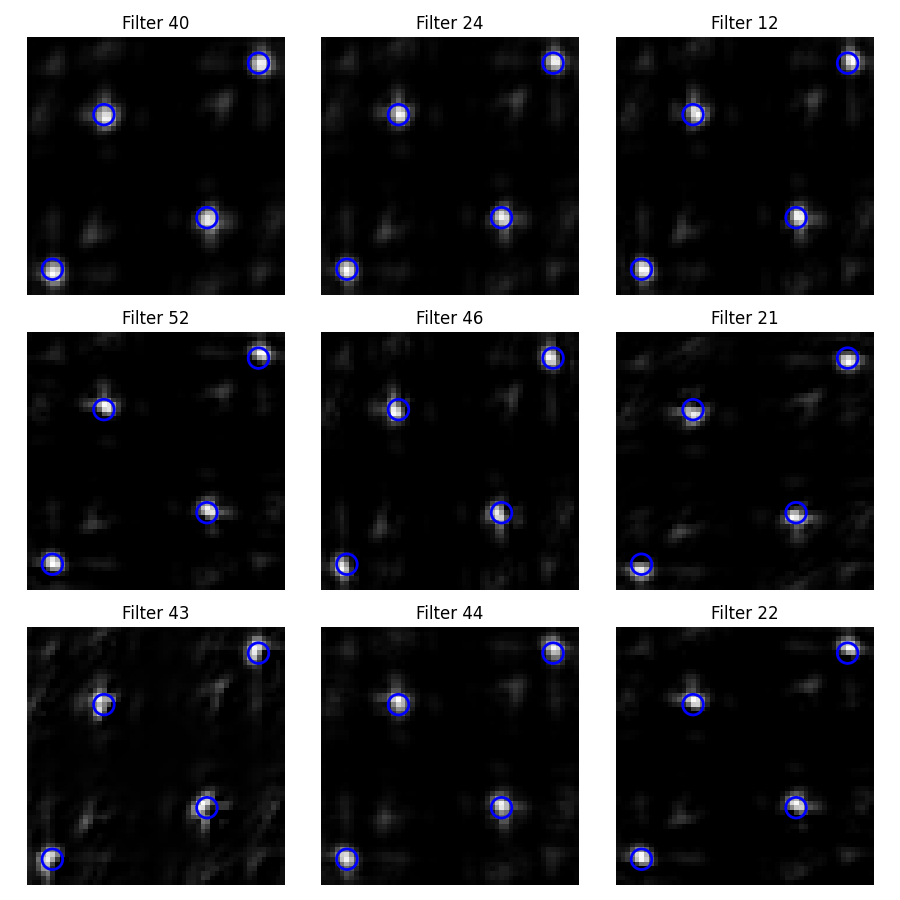}
\caption{Effect of the top 9 activating convolutional filters in the first layer for the cases of Fig. \ref{fig10}. In the right panels (localized example), the short PO scar identified by our NN is marked by means of blue empty circles.}
\label{fig11}
\end{figure}

\begin{figure}[ht]
\centering
\includegraphics[width=0.45\textwidth]{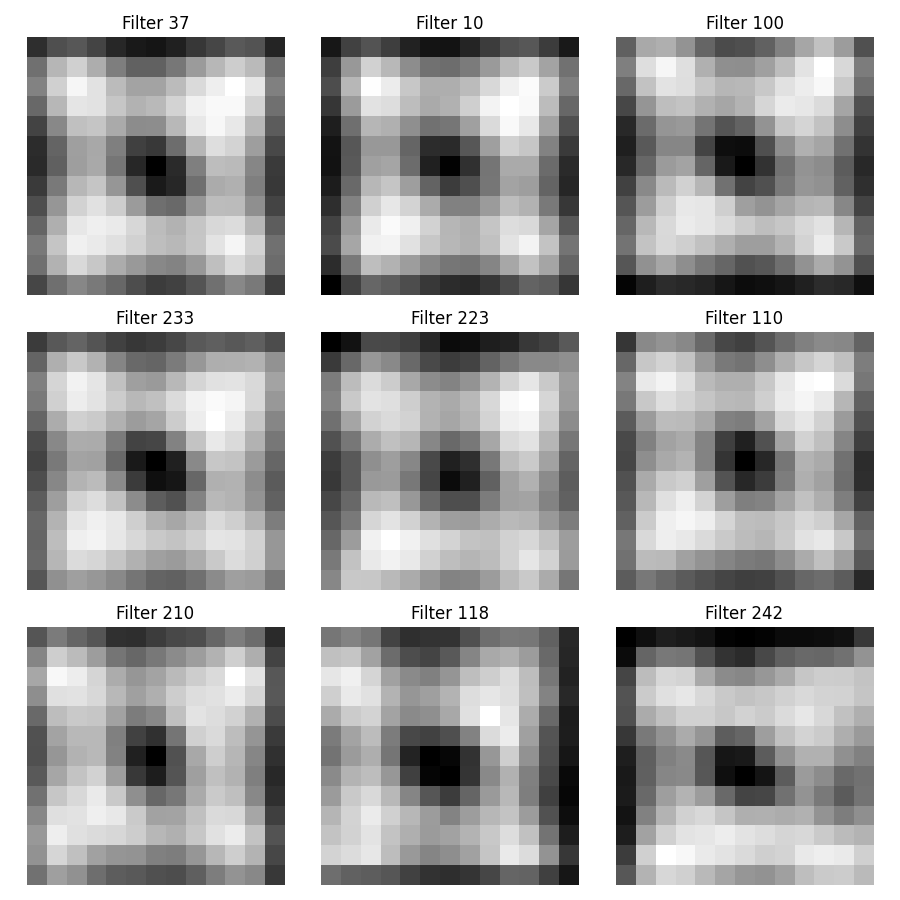}
\hfill
\includegraphics[width=0.45\textwidth]{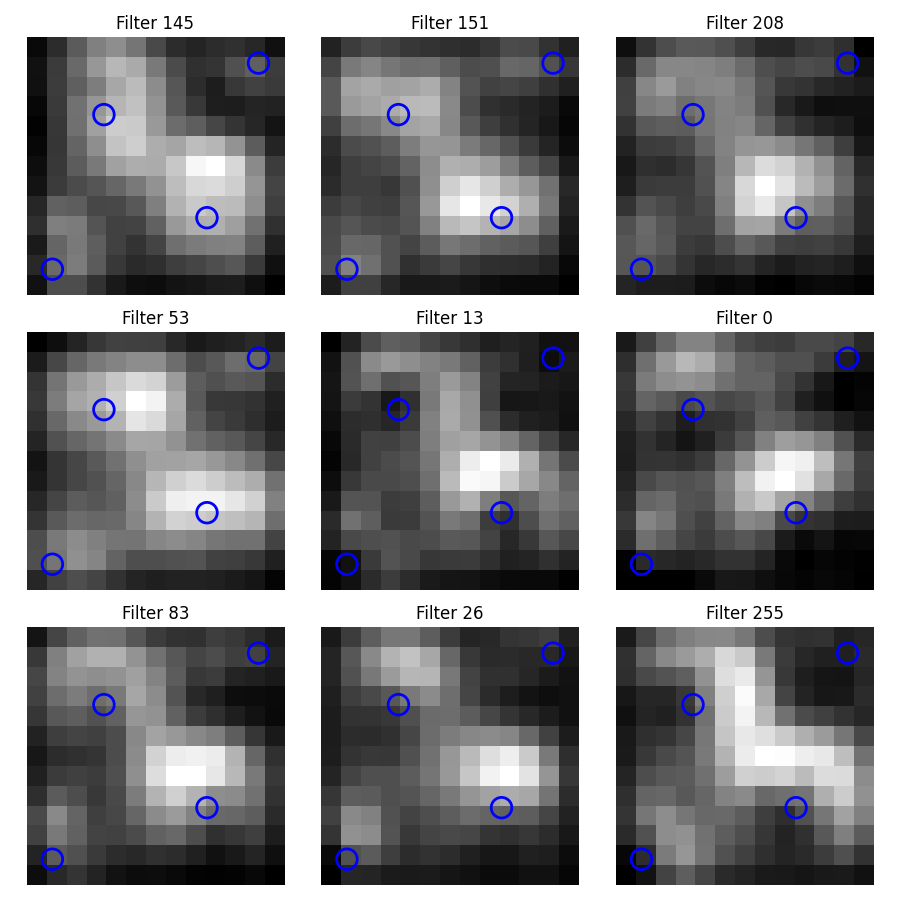}
\caption{Effect of the top 9 activating convolutional filters in the last layer for the cases of Fig. \ref{fig10}. In the right panels (localized example), the short PO scar identified by our NN is marked by means of blue empty circles.}
\label{fig12}
\end{figure}

The same behaviour has been found for the partially open map, which we explain in the following. 
In Fig. \ref{fig9} we show the LR representations corresponding to 9 randomly selected wave functions. 
We have picked the 7th and 9th resonances (see Fig. \ref{fig10}) as examples of non-localization and localization, respectively. 
In the last layer, the action of the filters wipes out the features of the non-localized resonance inside the principal area supporting the repeller, strikingly hinting where the main opening is located (its first iteration). In the localized example the same behavior as in the closed case is found (see Figs. \ref{fig11} and \ref{fig12}). In fact, a short PO of period 4 is singled out at all levels of the layers. 

\section{Conclusions}
 \label{sec:Conclusions}

We have defined a simple network architecture consisting of a custom "quantum" NN first part, integrated with a modified AlexNet. This allowed us to efficiently classify resonances of the paradigmatic open quantum tribaker map into localized and non-localized. This was achieved with independence of the system size, a fact that demonstrates the scalability of our model in this effective proof of principle. 

When analyzing the way in which the NN learns to distinguish localized from non-localized resonances, we have found that the process is able to detect POs scarring and the main features of the repellers. The physical meaning of the custom "quantum" NN in the first part of our model design is crucial not only for allowing this interpretations of the learning process, but also for making the ML affordable in terms of computational cost. The coherent states adapt to the most efficient distribution in order to classify the resonances in terms of their localization.

The detailed performance analysis has shown that our model not only works very well in the 
closed scenario, perhaps it is even more suitable for the partially open case. But this tells us more than just that; indeed, we have a hint towards a better understanding of localization in partially open systems thanks to ML. The ML performance measures seem to be almost localization morphology indicators by themselves, suggesting a sharper separation between localized and non-localized resonances which actually helps the NN to learn faster and better.

Through our ML integrated model we were able to reach an affordable method for systematically classifying the morphology of generic wave functions. This affordability is essential for the ultimate goal of exploring the localization rate going deep into the semiclassical limit. This is a way to overcome the last barrier into answering a long standing question in the quantum scattering field: is there more, the same, or less localization in (partially) open systems than in closed ones? \cite{Novaes}. We are now pursuing that objective by means of extensive calculations.

\section{Acknowledgments}
Support from CONICET is gratefully acknowledged.
This research has also been partially funded by the Spanish Ministry of Science, 
Innovation and Universities, Gobierno de Espa\~na under Contract No.~PID 2021-122711NB-C21.

%

\begin{thebibliography}{00}

\bibitem{Blumel1} 
R. Blümel and U. Smilansky,
Phys. Rev. Lett. {\bf 60}, 477 (1988).

\bibitem{Blumel2} 
R. Blümel and U. Smilansky,
Phys. Rev. Lett. {\bf 64}, 241 (1990).

\bibitem{Mendez} 
R.A. Méndez-Sánchez, U. Kuhl, M. Barth, C.H. Lewenkopf, and H.-J. Stöckmann,
Phys. Rev. Lett. {\bf 91}, 174102 (2003).

\bibitem{art0} 
D. Wisniacki and G.G. Carlo,
Phys. Rev. E {\bf 77}, 045201(R) (2008).

\bibitem{Dietz} 
B. Dietz, T. Friedrich, H.L. Harney, M. Miski-Oglu, A. Richter, F. Schäfer, 
J. Verbaarschot, and H.A. Weidenmüller,
Phys. Rev. Lett. {\bf 103}, 064101 (2009).

\bibitem{Novaes} 
M. Novaes, 
J. Phys. A: Math. Theor. {\bf 46}, 143001 (2013).

\bibitem{Cao}
H. Cao and J. Wiersig, 
Rev. Mod. Phys. {\bf 87}, 61 (2015).

\bibitem{RecentExp} 
S. Bittner, K. Kim, Y. Zeng, Q.J. Wang, and H. Cao, 
New J. Phys. {\bf 22}, 083002 (2020).

\bibitem{HusimisNew}
J. Hall, S. Malzard, E.-M. Graefe, 
Phys. Rev. Lett. {\bf 131}, 040402 (2023).

\bibitem{Ketz1}
K. Clauß, M.J. Körber, A. Bäcker, and R. Ketzmerick,
Phys. Rev. Lett. {\bf 121}, 074101 (2018).

\bibitem{Ketz2}
K. Clauß, E.G. Altmann, A. Bäcker, and R. Ketzmerick,
Phys. Rev. E {\bf 100}, 052205 (2019).

\bibitem{Ketz3}
K. Clauß, F. Kunzmann, A. Bäcker, and R. Ketzmerick,
Phys. Rev. E {\bf 103}, 042204 (2021).

\bibitem{Ketz4}
K. Clauß and R. Ketzmerick,
J. Phys. A: Math. Theor. {\bf 55} 204006 (2022).

\bibitem{Ketz5}
R. Ketzmerick, K. Clauß, F. Fritzsch, and A. Bäcker, 
Phys. Rev. Lett. {\bf 129}, 193901 (2022).

\bibitem{Ketz6}
J.R. Schmidt and R. Ketzmerick, 
New J. Phys. {\bf 25} 123034 (2023).

\bibitem{our1}
G.G. Carlo, R.M. Benito, and F. Borondo, 
Phys. Rev. E {\bf 94}, 012222 (2016).

\bibitem{our2}
C.A. Prado, G.G. Carlo, R.M. Benito, and F. Borondo,
Phys. Rev. E {\bf 97}, 042211 (2018).

\bibitem{our3}
J. Montes, G.G. Carlo, and F. Borondo,
Commun. Nonlinear Sci. Numer. Simulat. {\bf 132}, 107886 (2024). 

\bibitem{ML1}
G. Carleo, I. Cirac, K. Cranmer, L. Daudet, M. Schuld,
N. Tishby, L. Vogt-Maranto, and L. Zdeborová, 
Rev. Mod. Phys. {\bf 91}, 045002 (2019).

\bibitem{ML2}
N. Asif, U. Khalid, A. Khan, T.Q. Duong, and H. Shin, 
Sci. Rep. {\bf 13}, 1562 (2023).

\bibitem{ML3}
L. Domingo, G. Carlo, and F. Borondo, 
Sci. Rep. {\bf 13}, 8790 (2023).

\bibitem{ML4}
L. Domingo, G. Carlo, and F. Borondo, 
Phys. Rev. E {\bf 106}, L043301 (2022).

\bibitem{AlexNet}
A. Krizhevsky, I. Sutskever, and G.E. Hinton, 
Communications of the ACM {\bf 60}, 84 (2017). 

\bibitem{art1} 
M. Novaes, J.M. Pedrosa, D. Wisniacki, G.G. Carlo, and J.P. Keating, 
Phys. Rev. E {\bf 80}, 035202(R) 2009.

\bibitem{art2} 
L. Ermann, G.G. Carlo, and M. Saraceno, 
Phys. Rev. Lett. {\bf 103}, 054102 (2009).

\bibitem{art3} 
J.M. Pedrosa, D. Wisniacki, G.G. Carlo, and M. Novaes, 
Phys. Rev. E {\bf 85}, 036203 (2012).

\bibitem{art4} 
L. Ermann, G.G. Carlo, J.M. Pedrosa, and M. Saraceno, 
Phys. Rev. E {\bf 85}, 066204 (2012).

\bibitem{art5} 
G.G. Carlo, D.A. Wisniacki, L. Ermann, R.M. Benito, and F. Borondo, 
Phys. Rev. E {\bf 87}, 012909 (2013).

\bibitem{Ozorio 1994}
M. Basilio De Matos, A.M. Ozorio De Almeida, Ann. Phys. {\bf 237}, 46-65 (1995).

\bibitem{Hannay 1980}
J.H. Hannay, M.V. Berry, Physica D 1 267 (1980).

\bibitem{Espositi 2005}
M. Degli Espositi, B. Winn, J.Phys.A: Math.Gen.{\bf 38}, 5895-5912 (2005).

\bibitem{Saraceno1}
M. Saraceno, Ann. Phys. \textbf{199}, 37 (1990); M. Saraceno and R.O. Vallejos,

Chaos \textbf{6}, 193 (1996); A. \L ozi\'{n}ski, P. Pako\'{n}ski and K. \.{Z}yczkowski,
Phys. Rev. E \textbf{66}, 065201(R) (2002).

\bibitem{Saraceno2}
M. Saraceno and A. Voros, Physica D \textbf{79}, 206 (1994).

\bibitem{Shnirelman}
A.I. Shnirelman (in Russian), Usp. Math. Nauk {\bf 29}, 181 (1974).

\bibitem{Vergini}
E.G. Vergini, Phys. Rev. Lett. {\bf 108}, 264101 (2012); 
E.G. Vergini, EPL {\bf 110}, 10010 (2015).



%
\end{thebibliography}
\end{document}